\documentclass[aps,prl,twocolumn,amssymb,superscriptaddress,floatfix]{revtex4-2}

\usepackage{graphicx}
\usepackage{siunitx}
\usepackage{color}
\usepackage[hidelinks]{hyperref}

\newcommand{\costh}{\ensuremath{\langle \cos^2 \theta_{\text{2D}} \rangle}}
\newcommand{\notwo}[0]{$\mathrm{(NO)_2}$}

\begin{document}

\title{Control of molecular rotation in helium nanodroplets with an optical centrifuge}

\author{Ian MacPhail-Bartley}

\author{Alexander A. Milner}
\affiliation{Department of Physics \& Astronomy, The University of British Columbia, Vancouver, Canada}

\author{Frank Stienkemeier}
\affiliation{Institute of Physics, University of Freiburg, Freiburg, Germany}

\author{Valery Milner}
\email[]{vmilner@phas.ubc.ca}
\affiliation{Department of Physics \& Astronomy, The University of British Columbia, Vancouver, Canada}

\date{\today}

\begin{abstract}
We experimentally demonstrate that the rotation of molecules embedded in helium nanodroplets can be controlled with an optical centrifuge, allowing for the study of molecular dynamics inside the strongly interacting many-body environment of superfluid helium at variable levels of rotational excitation. By doping the droplets with dimers of nitric oxide, \notwo{}, and measuring the degree of their centrifuge-induced alignment as a function of time, we show both the forced in-field rotation of molecules in a continuous range of frequencies, as well as the field-free resonant rotation with a long nanosecond-scale decay. The ability to control and monitor the rotational dynamics of molecular rotors inside the superfluid medium may shed new light on superfluidity and the interaction of superfluids with defects at the atomic level.
\end{abstract}

\maketitle
Molecules solvated in, and interacting with superfluid helium can serve as ``nanoprobes'' of superfluidity, which is not yet fully understood at the microscopic level~\cite{Molecules-in-Superfluid-Helium2022,Superfluid-Helium-Droplets:2004,Infrared-spectroscopy-of-HOCl2012,Critical-Landau-Velocity2013,Dissipative-vibrational-wave-packet-dynamics2010,Dynamics-of-molecular-rotors2023,Superfluidity-Within-a-Small1998,Photoionisaton-of-pure-and-doped2014}. Infrared spectroscopy of molecules embedded inside helium nanodroplets has demonstrated that some molecules rotate freely, albeit with renormalized rotational and centrifugal distortion constants~\cite{Superfluid-Hydrodynamic-Model1999,Rotation-in-liquid-He4-Lessons2001}, while others strongly couple to the surrounding helium environment, as indicated by a few-orders-of-magnitude increase of their spectral line widths~\cite{Superfluid-Helium-Droplets:2004}. The degree of this coupling appears to depend on whether the rotational energy of a molecule matches the energy of elementary collective modes of the superfluid~\cite{Roton-Rotation-Coupling-of-Acetylene2004,Infrared-spectroscopy-of-helium2006,A-simple-model-for-high2022,Infrared-spectroscopy-of-HOCl2012}. The most stark differences in rotational line widths, when compared to those of the same molecule in the gas phase, were found in the energy range of a roton peak around \SI{180}{GHz}, corresponding to the high density of states in the superfluid's excitation spectrum~\cite{Infrared-spectroscopy-of-helium2006,Specific-heat-and-dispersion1981}.

Although powerful in examining the effects of a host medium on the rotational spectrum of the embedded molecule, infrared spectroscopy is limited to the lowest rotational states populated at the low temperature of a helium nanodroplet ($\approx\SI{0.4}{K}$). Hence, studying the frequency dependence of the molecule-superfluid coupling in the energy region of interest may be accomplished only by means of choosing multiple molecules with different rotational constants~\cite{Infrared-spectroscopy-of-helium2006}. However, this is often accompanied by other changes in the molecular structure. For instance, molecules with higher degrees of symmetry were found to exhibit narrow rotational lines, despite the fact that their rotational energy was comparable to the energy of collective excitations of the helium bath~\cite{Ravi2011}. Together with the constraints of being applicable to IR-active transitions only, the outlined limitations of the spectroscopic approach motivate the search for alternative methods of investigating superfluidity with molecular rotors, which would allow to cover broad rotational energy range while using a single molecular species.

Complementary to the frequency-resolved rotational spectroscopy, time-resolved techniques have been developed and used to study the rotation of molecules embedded inside~\cite{Pentlehner2013, Rotational-Coherence-Spectroscopy2020,Excited-rotational-states2021, Femtosecond-Rotational-Dynamics2022} and on the surface~\cite{Nonadiabatic-laser-induced-alignment2025} of helium nanodroplets, as well as in bulk superfluid helium~\cite{Dynamics-of-molecular-rotors2023,Coherent-control-of-molecular2024}. Since these methods are based on a coherent evolution of the laser-induced rotational wave packet, they have been successful in investigating rotational decoherence in the helium environment. In these experiments, however, the decoherence timescale cannot be distinguished from the timescale of the rotational energy relaxation, making it difficult to study the two fundamentally different phenomena separately. Equally complicated is the analysis of the frequency dependence of the observed decay due to the broad distribution of rotational states in, and correspondingly high frequency content of, the impulsively excited wave packets.

\begin{figure*}
\includegraphics[width=1.9\columnwidth]{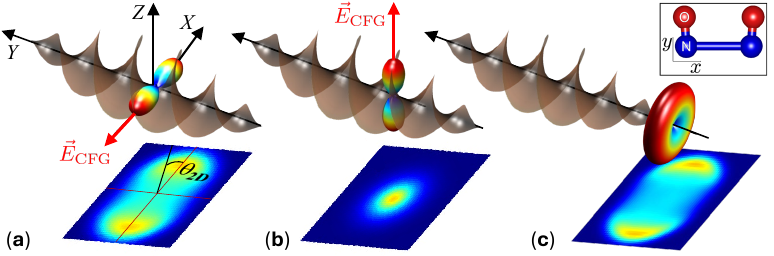}
\caption{\label{fig:Geometry} Geometry of the experiment. The brown corkscrews represent the electric field of the centrifuge, $\mathbf{\vec{E}}_{\mathrm{CFG}}$, propagating along the $Y$ axis and rotating in the $XZ$ plane. The TOFVMI setup is oriented along the $Z$ axis, projecting a 2D velocity map image of ion fragments onto the $XY$ plane. $\theta_{\text{2D}}$ in panel (\textbf{a}) is the angle of the detected ion hits with respect to the $X$ axis. Two red ``dumbbells'' and a ``donut'' illustrate the distributions of the \mbox{N-N} axis of the NO dimer (inset in the upper right corner) in three representative situations: alignment to the centrifuge field \textit{during} the excitation process, when $\mathbf{\vec{E}}_{\mathrm{CFG}}$ is either parallel (\textbf{a}) or perpendicular (\textbf{b}) to the detector plane, and confinement to the plane of rotation \textit{after} the excitation by the centrifuge (\textbf{c}). Numerically simulated VMI images are shown below the corresponding distributions.}
\end{figure*}
In this work, we employ the method of an ``optical centrifuge''~\cite{Optical-Centrifuge-for-Molecules1999,Forced-Molecular-Rotation2000} to spin molecules, embedded in He nanodroplets, in a controlled way. This enables us to (A) establish molecular rotation in a continuous range of frequencies, regardless of the molecule's rotational constant, and (B) directly observe the rotational energy decay, even in the presence of rapid rotational decoherence. Previous attempts to utilize this approach were not successful due to the fast rotational acceleration of the conventional optical centrifuge, and its high terminal frequency~\cite{Building-an-Optical-Centrifuge:2018,Control-of-molecular-rotation2024}, well beyond the upper limit set by the extremely high effective centrifugal distortion constant of the dopant molecule, interacting with the helium bath~\cite{Excited-rotational-states2021}. Here, we use a newly-developed ``constant-frequency centrifuge'' (cfCFG), which exhibits zero acceleration and can be tuned within the low frequency window below \SI{100}{GHz}~\cite{An-ultra-slow-optical-centrifuge.2025}. We demonstrate that cfCFG is well suited for studying molecular rotation in He nanodroplets across the energy scale comparable to the elementary excitations of superfluid helium.

An optical centrifuge is a linearly polarized laser pulse, whose polarization vector rotates around the laser propagation axis. Owing to the interaction between the field of the centrifuge and the induced dipole moment in the molecule of interest, the latter aligns with respect to the field polarization and follows its rotation, as long as the angular acceleration is not too fast~\cite{Adiabatic-excitation-of-rotational2004}. The conventional centrifuge rotates with accelerations on the order of \SI{100}{GHz/ps}, producing narrow rotational wave packets with central frequencies on the scale of a few THz~\cite{Laser-control-of-molecular2020}. In contrast, the constant-frequency centrifuge, implemented in this work, rotates with much lower rotational frequencies and zero acceleration~\cite{An-ultra-slow-optical-centrifuge.2025}. Unlike the conventional design, based on the femtosecond pulse shaper~\cite{Forced-Molecular-Rotation2000}, cfCFG is formed by splitting a frequency-chirped pulse in a Michelson interferometer. The chirp rate and time delay between the two arms of the interferometer determines the instantaneous frequency difference between the two pulses. When combined with opposite senses of circular polarization, their interference results in the linearly polarized field rotating with a constant frequency $f_{\text{CFG}}$, equal to half the instantaneous frequency difference between the two pulses. The cfCFG used for these experiments was created from the uncompressed output of a chirped pulse amplifier (Spectra-Physics Spitfire Ace), which produces pulses stretched to \SI{\approx330}{\pico \second} (full width at half maximum, FWHM), with a bandwidth of \SI{\approx9}{\nano \metre} (FWHM), centered at \SI{800}{\nano \metre}. The centrifuge retains the approximately Gaussian intensity profile of the laser output.

We use the conventional method of producing a beam of helium nanodroplets, doped with the molecule of interest (for a recent review of this common technique, see Ref.~\citenum{Molecules-in-Superfluid-Helium2022}). In our experiments, helium droplets are formed by expanding $30$~bar of high-purity helium through a $5$~$\mu$m orifice, cryogenically cooled to $18$~K, inside a vacuum chamber. Under these conditions, the droplets on average contain about $3000$~He atoms~\cite{Superfluid-Helium-Droplets:2004}. The nanodroplet beam passes through a pick-up cell, where the partial pressure of the dopant molecule is adjusted using a leak valve. The latter allows one to maximize the number of droplets with the target molecular species embedded within, while ensuring that a negligible number of droplets contain larger clusters of the dopant~\cite{Successive-capture-and-coagulation1995} (for more technical details on our droplet apparatus and doping technique, see Ref.~\citenum{Control-of-molecular-rotation2024}).

\begin{figure}
\includegraphics[width=0.8\columnwidth]{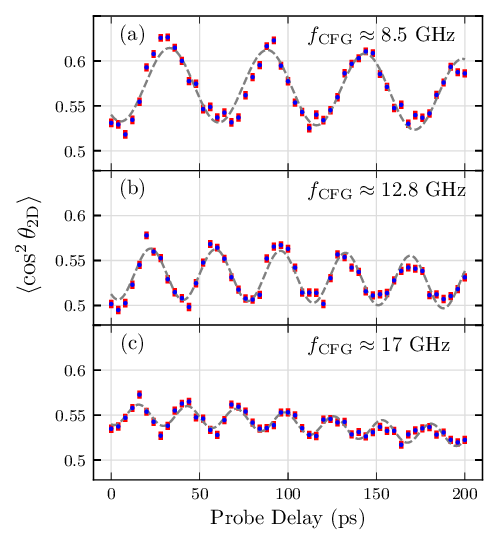}
\caption{\label{fig:ForcedRotation} Centrifuge-induced rotation of \notwo{} molecules in helium nanodroplets. $\costh(t)$ (blue markers with red error bars) was recorded during the centrifuge pulse, whose constant rotation frequency $f_{\text{CFG}}$ was tuned to (a) \SI{\approx8.5}{GHz}, (b) \SI{\approx13}{GHz}, and (c) \SI{\approx17}{GHz}. Values of \costh{} at the peaks of the oscillations indicate molecular alignment along the laboratory $X$ axis, while values at the troughs represent molecular alignment along the $Z$ axis (see Fig.~\ref{fig:Geometry}). A probe delay of $t=0$ corresponds to the peak of the cfCFG intensity profile. Grey dashed lines are fits to decaying sinusoids with frequencies of (a) \SI{18.1(3)}{GHz}, (b) \SI{27.2(3)}{GHz}, and (c) \SI{36.2(5)}{GHz}, respectively, closely matching the anticipated values of $2\times f_{\text{CFG}}$.}
\end{figure}
The doped helium droplet beam enters a time of flight velocity map imaging (TOFVMI) spectrometer~\cite{Velocity-map-imaging-of-ions1997} in the detection chamber, where it is crossed perpendicularly by the cfCFG laser beam. A short ($\approx120$~fs) probe pulse, collinear with the cfCFG beam and linearly polarized parallel to the TOFVMI axis, Coulomb explodes the dopant molecules. The cfCFG and probe pulses are focused onto the helium droplet beam by a lens (focal length $f=20$~cm), resulting in peak intensities of $\approx2\times10^{12}$~W$/$cm$^{2}$ and $\approx5\times10^{14}$~W$/$cm$^{2}$, respectively. After the Coulomb explosion, the distribution of molecular axes is inferred from the angular distribution of ion fragments detected by TOFVMI, using the conventional metric of \costh{}~\cite{Aligning-molecules-with1999}. Here, $\theta_{\text{2D}}$ is the angle of the ion fragment's velocity with respect to the rotational plane of the centrifuge (illustrated in the bottom left VMI image of Fig.~\ref{fig:Geometry}), and $\langle ... \rangle$ refers to averaging over the detected angles of several thousand ions.

We doped helium droplets with the dimers of nitric oxide, \notwo{}, as our molecule of choice for two reasons. First, because \notwo{} is not stable at room temperature, the background signal from gas-phase molecules, which can significantly alter \costh{}, is therefore negligible (as confirmed by blocking the droplets with an internal gate valve). Second, the utility of the TOFVMI technique in measuring the spatial distribution of molecular axes is known to deteriorate rapidly with the decreasing mass of ion fragments due to their scattering from He atoms during the Coulomb explosion process~\cite{Deconvoluting-nonaxial-recoil2016}. NO$^{+}$ fragments prove sufficiently heavy to minimize this effect and retain the angular information about the parent NO dimer at the time of explosion.

The planar ONNO geometry depicted in the inset of Fig.~\ref{fig:Geometry} is the most stable isomer of \notwo{} (acyclic cis-ONNO~\cite{High-level-theoretical-study2012, Relative-stabilities-of-NO22001}), and the most prevalent isomer formed in helium droplets~\cite{Infrared-Spectroscopy-and-Structure2016}. The most polarizable axis of the dimer is the axis connecting the two nitrogen atoms (\mbox{N-N} axis)~\cite{NIST-Computational-Chemistry2022}. Hence, we expect this axis to align with the polarization vector of the centrifuge, and follow its rotation in the $XZ$ plane. The time-dependent behavior of the centrifuge-induced molecular alignment can then be determined by measuring \costh{} of the NO$^{+}$ fragments as a function of the cfCFG-probe delay.

The experimentally observed gas-phase rotational constants of \notwo{} are: $B_{x} = 0.86$~cm$^{-1}$, $B_{y} = 0.19$~cm$^{-1}$, and $B_{z} = 0.15$~cm$^{-1}$, and the centrifugal distortion constant is  $D\approx$~\SI{1E-6}{cm^{-1}}~\cite{The-v-1-band-of-NO21993,The-NO-Dimer1997}. When the latter is ignored, the rotational energies of the near-prolate \notwo{} can be approximated by $E(J,K) = B_{yz}J(J+1) + (B_x - B_{yz})K^2$, where $B_{yz} = (B_y + B_z)/2$ and $K = \pm~0,1,...,J$. Owing to the low temperature of the droplets, the dimers are primarily in their ground rotational state $(J,K=0)$. In the prolate limit, the interaction with the laser field conserves $K$, hence $E(J)\approx B_{yz}J(J+1)$, implying that the two lowest gas-phase rotational transition frequencies $c\Delta E(J=0 \rightarrow J=2)$ and $c\Delta E(J=1 \rightarrow J=3)$ (where $c$ is the speed of light) are $\approx 31$~GHz and $\approx 51$~GHz, respectively.

\begin{figure}
\includegraphics[width=0.8\columnwidth]{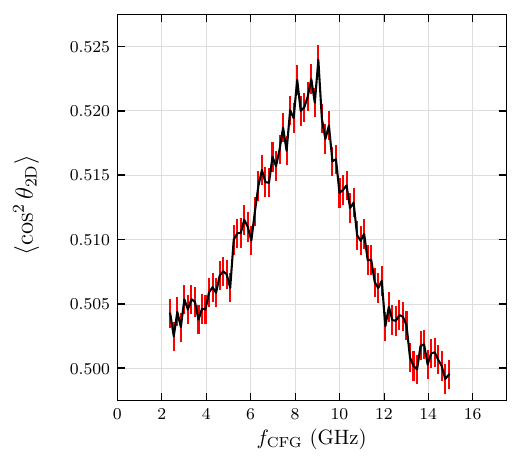}
\caption{\label{fig:FrequencyScan} \costh{} of \notwo{} in helium droplets, at a probe delay of \SI{550}{ps}, measured as a function of the rotational frequency of the optical centrifuge (black curve with red error bars). The latter exhibits a small degree of rotational acceleration (due to the third-order dispersion), amounting to the frequency difference of about \SI{4}{GHz} over its full duration~\cite{An-ultra-slow-optical-centrifuge.2025}. We consider this the primary factor contributing to the observed lineshape and its finite linewidth.}
\end{figure}

\begin{figure*}
\includegraphics[width=0.85\textwidth]{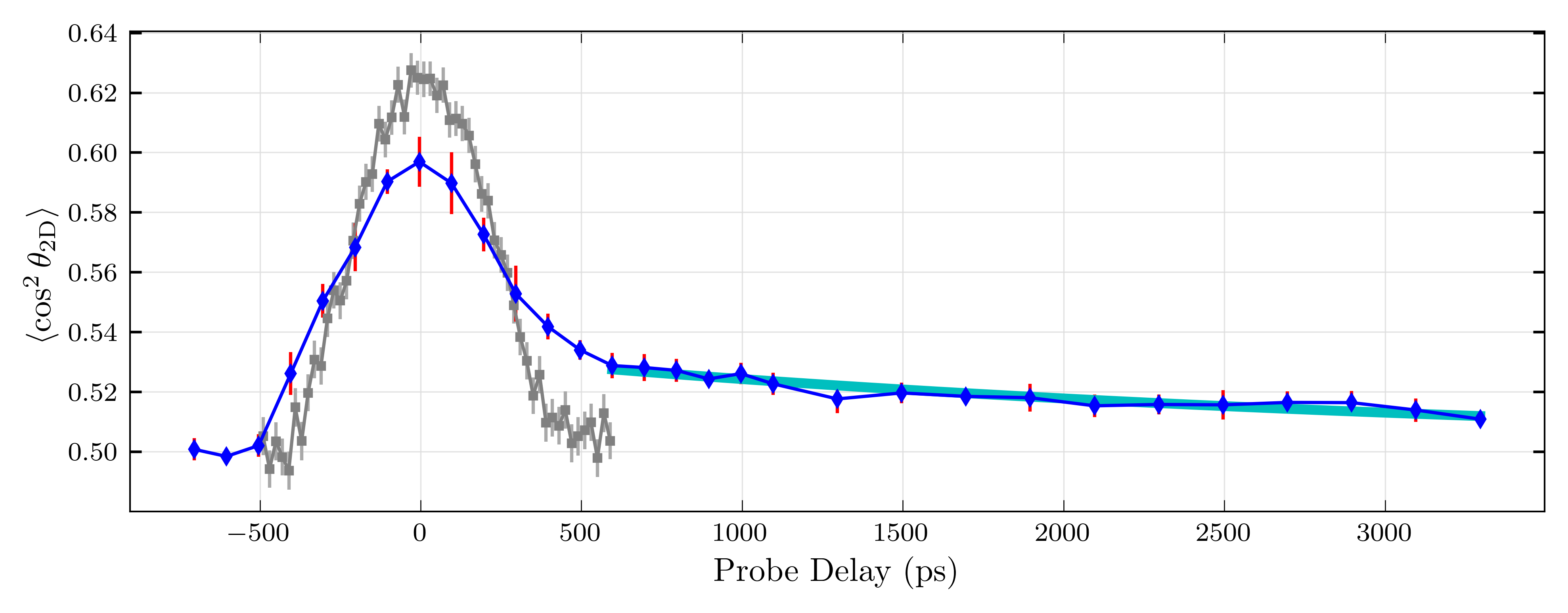}
\caption{\label{fig:LongScan} $\costh (t)$ of \notwo{} in helium droplets (blue diamonds with error bars), with cfCFG tuned to the resonant frequency of $f_{\text{CFG}} = 8.5$~GHz. The cyan curve represents an exponential fit (Eq.~\ref{eq:fit}) of the field-free decay in $\costh(t)$ after the centrifuge, with a fixed asymptote of $\costh(t\rightarrow \infty) = 0.5$. The extracted decay constant is $\tau=3200$~ps~$\pm~300$~ps. Also plotted (grey squares with error bars) is the adiabatic alignment~\cite{Alignment-and-Trapping-of-Molecules1995} of \notwo{} by a pulse of the same duration as cfCFG, linearly polarized parallel to the plane of the TOFVMI detector.}
\end{figure*}
We start by exploring the behavior of \notwo{} molecules during their exposure to the centrifuge pulse, hereafter referred to as the ``in-field'' dynamics. Fig.~\ref{fig:ForcedRotation} shows the observed value of \costh{} as a function of time, starting from the peak of the centrifuge at $t=0$. $\costh (t)$ was measured with the rotation frequency $f_{\text{CFG}}$ set to three different values around \SI{8.5}{GHz}, \SI{13}{GHz} and \SI{17}{GHz}, based on the estimated cfCFG parameters found in our earlier work~\cite{An-ultra-slow-optical-centrifuge.2025}.

Clear oscillations at $\approx 2 \times f_\text{CFG}$ confirm that NO dimers are following the rotation of the linearly polarized centrifuge field. Indeed, as the molecules rotate in the $XZ$ plane, the value of \costh{} alternates between higher and lower values, corresponding to the average molecular alignment parallel and perpendicular to the plane of the TOFVMI detector, as illustrated by the dumbbell-shape distributions in Figs.~\ref{fig:Geometry}(\textbf{a}) and \ref{fig:Geometry}(\textbf{b}), respectively. Similarly to the well know adiabatic molecular alignment by a non-rotating intense linearly polarized laser pulse, the rotating alignment observed here stems from the field-induced pendular states, i.e. bare rotational states dressed by the centrifuge field~\cite{Alignment-and-Trapping-of-Molecules1995}.

Multiple factors contribute to the exact values of \costh{}, with the maximum degree of alignment primarily dictated by the molecule's polarizability anisotropy, and the intensity of the laser field~\cite{Colloquium:-Aligning-molecules2003}. The decreasing amplitude of oscillations with increasing $f_{\text{CFG}}$ indicates that fewer molecules are captured by the centrifuge, when the deviation of its constant frequency from the initial (zero) frequency of molecules in the ground rotational state becomes larger.

The amplitude of the observed in-field oscillations, described above, follows the intensity envelope of the centrifuge field, with the value of \costh{} approaching its isotropic value of 0.5 by the end of the centrifuge pulse. This behavior is analogous to adiabatic molecular alignment by a long linearly polarized laser field~\cite{Alignment-and-Trapping-of-Molecules1995, Colloquium:-Aligning-molecules2003}. In both cases, since the two-photon spectrum of the aligning field does not contain any rotational frequencies of the molecule, the latter is transferred back to its initial isotropic state as the field turns off adiabatically.

On the other hand, when $f_{\text{CFG}}$ is resonant with one of the molecule's intrinsic rotational frequencies, the resonant two-photon Raman process transfers population to the excited rotational state, creating a rotational wave packet which outlasts the excitation pulse~\cite{Adiabatic-excitation-of-rotational2004}. Since the frequency difference between the two circularly polarized centrifuge arms is equal to $2 f_{\text{CFG}}$~\cite{An-ultra-slow-optical-centrifuge.2025}, this occurs when $2 f_{\text{CFG}} = c\Delta E(J \rightarrow J+2)$. To find these resonances, corresponding to long-lived field-free rotation, we measured $\costh$ at a probe delay of \SI{550}{ps}, well beyond the end of cfCFG, while scanning $f_{\text{CFG}}$. The result is shown in Fig.~\ref{fig:FrequencyScan}.

As \notwo{} is primarily in its ground rotational state at $0.4$~K, the constant-frequency centrifuge can only drive a single Raman transition from $J=0$ to $J=2$, to which we assign the observed peak at $f_{\text{CFG}}=\SI{8.4(2)}{GHz}$. From this frequency, we extract the value of $B_{yz} = \SI{0.092(2)}{cm^{-1}}$,  which is a factor of $1.9$ smaller than the corresponding gas-phase value. This finding is consistent with the typical  renormalization of rotational constants reported for many other molecules in helium droplets~\cite{Infrared-spectroscopy-of-helium2006}.

Using our ability to create a rotational wave packet at the resonant frequency found from Fig.~\ref{fig:FrequencyScan}, we investigated its field-free evolution by recording \costh{} for probe delays up to \SI{3.3}{ns}, as depicted in Fig.~\ref{fig:LongScan}. We note that despite rotational dephasing, manifested by the disappearance of oscillations in the time dependence of \costh{}, centrifuged molecules undergoing unidirectional rotation maintain the alignment of their axes in the plane of rotation. The latter is illustrated by the donut shape in Fig.~\ref{fig:Geometry}(\textbf{c}), and by \costh$> 0.5$, which reflects an anisotropic axis distribution and allows the detection of the rotational energy decay. In this regard, our results are distinct from, and complementary to, the previous studies of molecular rotation in helium nanodroplets, based on the detection of rotational coherence~\cite{Rotational-Coherence-Spectroscopy2020}.

As can be seen in Fig.~\ref{fig:LongScan}, $\costh(t)$ increases to $\approx0.6$ during the centrifuge pulse (resembling its intensity profile), and then decays down to a $\costh \approx 0.52$, where the rotation persists for at least three nanoseconds. The behavior is contrasted with a non-resonant adiabatic alignment of NO dimers by a linearly polarized pulse of the same duration, which returns molecules to their initial non-rotating state (grey curve). Fitting the decaying part of the observed resonant signal after the centrifuge by a single exponent, and assuming the asymptotic value of $\costh{}=0.5$,
\begin{equation}
    \label{eq:fit}
    S(t) = 0.5 + A \exp{(-t/\tau)},
\end{equation}
\noindent we find the decay constant $\tau=3200$~ps~$\pm~300$~ps. This long decay is consistent with the rotational line widths in infrared spectra of molecules with similar rotational energies~\cite{Infrared-spectroscopy-of-helium2006}, and is attributed to the fact that those energies are well below the energy of collective excitations in the surrounding helium bath.

In summary, we experimentally demonstrated control of molecular rotation in superfluid helium using an optical centrifuge. This includes forcing in-field rotation in a continuous range of frequencies, as well as exciting a single field-free rotational resonance. Our preliminary observations suggest that the molecules follow the centrifuge field as long as its rotation frequency does not exceed that of the highest thermally populated rotational state. Quantitative analysis of this result will require further investigation. In the case of the long-lasting field-free rotation, we showed that its lifetime can be measured directly in the time domain, providing an alternative to the line width analysis in infrared spectroscopy.

Rotating a molecule with a constant well-defined angular velocity presents a new opportunity to investigate the response of the superfluid to a moving neutral defect at the atomic scale, complimenting previous studies with ions~\cite{Reif1960,Nancolas1985,Gunther1996}, large clusters~\cite{Buelna2016}, molecules in metastable Rydberg~\cite{Critical-Landau-Velocity2013,Dynamics-of-molecular-rotors2023,Coherent-control-of-molecular2024} and excited vibrational states~\cite{Dissipative-vibrational-wave-packet-dynamics2010}. Finally, scanning the rotational frequency of the optical centrifuge offers a new approach to measuring the rotational constant of molecules, embedded in helium nanodroplets and not amenable to IR spectroscopy. Applying this method to \notwo{} enabled us to determine the effect of helium environment on the rotational constant of the dimer.

Work is currently underway to implement a new ``ultra-slow'' optical centrifuge, expected to drive sequential rotational excitations of molecules in helium droplets~\cite{An-ultra-slow-optical-centrifuge.2025}. Adding the ability to control the degree of excitation will allow the interplay between rotational energy and state lifetimes to be investigated on a state-by-state basis, all with the same molecule. Applying the optical centrifuge to molecules in superfluid helium could also be used in the studies of controlled creation of quantized vortices~\cite{Spectroscopy-and-dynamics-in-helium2006} by unidirectional stirring the droplet from within. Since the studies of vortices and shapes of spinning helium droplets by means of coherent diffraction imaging techniques~\cite{Imaging-Quantum-Vortices2019,Shapes-and-vorticities-of-superfluid2014} already play a prominent role, controlling vortices may provide new opportunities also in those investigations.

\begin{acknowledgments}
This research has been supported by the grants from CFI, BCKDF, and NSERC, and carried out under the auspices of the UBC Center for Research on Ultra-Cold Systems (CRUCS). The authors thank Dr.~J.~Bertrand for his dedicated assistance with maintaining the laser system.
\end{acknowledgments}


%

\end{document}